\documentclass{PoS}
\usepackage{cancel}
\usepackage{slashed}
\usepackage{amsfonts}                            
\usepackage{amsmath}
\usepackage{amssymb}
\usepackage{graphicx}

\newcommand{\alfive}{{\alpha_5}}
\newcommand{\CP}{{CP}}
\newcommand{\CPbar}{{\overline{CP}}}
\newcommand{\CPviol}{{\cancel{CP}}}

\newcommand{\Tr}{{\mathrm{Tr}}}

\newcommand{\alfivehat}{{\hat\alpha_5}}

\newcommand{\MSbar}{{\overline{MS}}}

\newcommand{\mcC}{{\cal C}}
\newcommand{\mcL}{{\mathcal L}}
\newcommand{\mcO}{{\mathcal O}}
\newcommand{\mcP}{{\cal P}}
\newcommand{\mcD}{{\mathcal D}}
\newcommand{\tsep}{t_{\mathrm{sep}}}
\newcommand{\tOp}{t_{\mathrm{op}}}
\newcommand{\vac}{{\mathrm{vac}}}

\title{Calculation of Nucleon Electric Dipole Moments 
  Induced by Quark Chromo-Electric Dipole Moments and the QCD $\theta$-term}

\ShortTitle{Neutron EDM induced by chromo-EDM and $\theta_\text{QCD}$}

\author{
  \speaker{Sergey~Syritsyn}${}^{a,b}$
  Taku Izubuchi${}^{b}$,
  Hiroshi~Ohki${}^{a,c}$,
  \\
  \llap{${}^a$} RIKEN/BNL Research Center, Brookhaven National Laboratory, 
      Upton, NY 11973, USA\\
  \llap{${}^b$} Department of Physics and Astronomy, Stony Brook University, 
      Stony Brook, NY 11794, USA \\
  \llap{${}^c$} Department of Physics, Nara Women's University, Nara 630-8506, Japan \\
E-mail: \email{ssyritsyn@quark.phy.bnl.gov}}

\abstract{
Electric dipole moments (EDMs) of nucleons and nuclei, which are sought as evidence of CP
violation, require lattice calculations to connect constraints from experiments to limits on the
strong CP violation within QCD or CP violation introduced by new physics from beyond the
standard model.
Nucleon EDM calculations on a lattice are notoriously hard due to large statistical noise,
chiral symmetry violating effects, and potential mixing of the EDM and the anomalous
magnetic moment of the nucleon.
In this report, details of ongoing lattice calculations of proton and neutron EDMs induced by
the QCD $\theta$-term and the quark chromo-EDM, the lowest-dimension effective CP-violating
quark-gluon interaction are presented.
Our calculation employs chiral-symmetric fermion discretization.
An assessment of feasibility of nucleon EDM calculations at the physical point is discussed.
}

\FullConference{XIII Quark Confinement and the Hadron Spectrum - Confinement2018\\
		31 July - 6 August 2018\\
		Maynooth University, Ireland}

\PoScopydate{\today}

\begin{document}
%

\section{Introduction}

Observing a non-zero nucleon electric dipole moment (nEDM) will be evidence of violation of
$T(\CP)$-symmetry beyond the level of the Standard Model (SM).
The Standard Model itself does not have sufficient magnitude of $\CP$ violation to explain the
observed excess of matter over antimatter in the Universe.
Knowledge of nucleon structure and interactions is required to translate precise EDM
measurements, which are projected to improve by two orders of magnitude in the next decade, into
constraints on $\CP$ violation at the quark-gluon level and bounds on the strong $\CP$-violation
($\theta_\text{QCD}$ angle) as well as beyond-the-Standard-Model (BSM) theories.
Such nucleon structure calculations are possible only with nonperturbative lattice QCD methods.
(Effective) interactions that can induce nucleon EDM have to be $P$- and $\CP$-odd, and can be
classified by their dimension~\cite{Engel:2013lsa}
\begin{equation}
\mcL^{\CPbar} = \sum_i \frac{c_i}{\Lambda_{(i)}^{d_i-4}} \mcO_i^{[d_i]}\,,
\end{equation}
where $d_i$ are the dimensions of the effective interaction densities $\mcO_i$ and
$\Lambda_{(i)}$ are the scales of the underlying, more fundamental interactions that cause them.
We use lattice QCD with chiral quark action to calculate the nEDM induced by 
the $d=4$ $\theta_\text{QCD}$-term,
as well as by the chromo-electric moment, the lowest-dimension ($d=5$) effective quark-gluon 
$\CP$-odd interaction that may be generated by extensions of the Standard Model:
\begin{align}
\nonumber
\sum_x \mcL^{\CPbar}_x &= i \theta_\text{QCD} Q + i \sum_x\sum_q \tilde\delta_q \mcC^q_x\,, 
\\
\label{eqn:Qtopo}
Q &= \frac1{16\pi^2} \sum_x\Tr \big[\hat{G}_{\mu\nu}\tilde{\hat{G}}_{\mu\nu}\big]_x\,,
\\
\label{eqn:cedm_def}
\mcC^q &= \bar{q}\,\big[\frac12 (\hat{G}_{\mu\nu}\sigma_{\mu\nu})\,\gamma_5\big]\, q\,,
\end{align}
where $\hat{G}_{\mu\nu} = g(G^{pert}_{\mu\nu})^a\lambda^a$ is the gluon field strength defined in
$su(3)$ algebra with generators satisfying $\Tr[\lambda^a\lambda^b]=\frac12\delta^{ab}$
and $\tilde{\hat{G}}_{\rho\sigma}=\frac12\epsilon_{\mu\nu\rho\sigma} \hat{G}_{\mu\nu}$.
On a lattice, this gluon field corresponds to a $1\times1$ plaquette
$U^P_{x,\mu\nu}\approx 1 + i a^2 \hat{G}_{x,\mu\nu} + O(a^4)$ 
but is typically approximated to higher order in the definition of $Q$.
In addition, we calculate nEDM induced by the pseudoscalar density 
\begin{equation}
\label{eqn:psc_def}
\mcP^q = \bar{q}\,\gamma_5 \, q\,,
\end{equation}
which are necessary for renormalizing the chromo-EDM operator $\mcC$.

\section{CP-odd nucleon structure on a lattice}

Nucleon electric dipole moments on a lattice can be calculated as the forward
limit of the $P,T$-odd electric dipole form factors
(EDFF) $F_3$~\cite{Shintani:2005xg,Berruto:2005hg,Aoki:2008gv,Guo:2015tla,Shindler:2015aqa,
Alexandrou:2015spa,Shintani:2015vsx,Abramczyk:2017oxr},
\begin{equation}
\label{eqn:ff_cpviol}
\langle p^\prime,\sigma^\prime |J^\mu|p,\sigma\rangle_{\CPviol}
  = \bar{u}_{p^\prime,\sigma^\prime} \big[
    F_1(Q^2) \gamma^\mu 
    + \big(F_2(Q^2) + iF_3(Q^2)\big) \frac{i\sigma^{\mu\nu}q_\nu}{2M_N}
  \big] u_{p,\sigma}\,,
\end{equation}
where $Q^2=-q^2$ and $q=p^\prime-p$.
Extrapolation $Q^2\to0$ is required to obtain the nEDM $d=F_3(0)$
because the $F_3$ contribution vanishes from Eq.~(\ref{eqn:ff_cpviol}) at $Q^2=0$.
$\CP$-odd interactions that induce nEDM are introduced as either fixed additional terms in
the lattice QCD action or as first-order perturbations to nucleon correlation functions.
In the latter  case, the nucleon-current correlators in $\CPviol$ vacuum are
\begin{gather}
S_\text{QCD}\to S_\text{QCD} + i \delta^\CPbar S = S_\text{QCD} + i\sum_{i,x} c_i [\mcO^{\CPbar}_i]_x\,,
\\
\label{eqn:corr_cpviol}
\begin{aligned}
\langle N\,[\bar q \gamma^\mu q]\, \bar N \rangle_{\CPviol}
&
= \frac1Z\int\,\mcD U\,\mcD\bar q\mcD q  e^{-S - i\delta^\CPbar S}
      \, N\,[\bar q \gamma^\mu q]\, \bar N 
\approx C_{NJ\bar N} - i\sum_i c_i \,\delta^\CPbar_i C_{NJ\bar N}
\,,
\end{aligned}
\end{gather}
where
$C_{NJ\bar N} = \langle N\,[\bar q \gamma^\mu q]\, \bar N\rangle$ and 
$\delta_i^\CPbar C_{NJ\bar N} = \langle N\,[\bar q \gamma^\mu q]\, \bar N \, 
  \sum_x [\mcO^\CPbar_i]_x \rangle$
are the nucleon-current correlation function and its $\CP$-odd perturbation, both evaluated 
in the usual $\CP$-even QCD vacuum.
To compute the matrix elements (\ref{eqn:ff_cpviol}) and extract the EDFF $F_3(Q^2)$, we
calculate the nucleon and nucleon-current correlators
\begin{align}
\label{eqn:twopt_cpviol}
\{\delta^{\CPbar}\}C_{N\bar N}(\vec p,t) 
  &= \sum_{\vec x} e^{-i\vec p\cdot \vec x}\langle N_{\vec x,t} \bar{N}_{\vec0, 0}
    \, \{\delta^{\CPbar}S\}\rangle_\CPviol\,,\\
\label{eqn:threept_cpviol}
\{\delta^{\CPbar}\} C_{NJ\bar N}(\vec p^\prime,\tsep;\vec q,\tOp)
  &= \sum_{\vec y,\vec z} e^{-i\vec p^\prime\cdot \vec y+i\vec q\cdot\vec z}
              \langle N_{\vec y,\tsep} J^\mu_{\vec z,\tOp} \bar{N}_{\vec0,0}
  \, \{\delta^{\CPbar}S\}\rangle\,.
\end{align}
with and without insertions of the $\CP$-odd interactions.
More details on the analysis of the form factors can be found in a recent
paper~\cite{Abramczyk:2017oxr}, where spurious contributions to nEDM due to
parity mixing $\langle vac |N|p\rangle \sim e^{i\alpha_5\gamma_5} u_p$
were discovered.

It has to be stressed that the correct treatment of the mixing and careful determination of the
mixing angle $\alpha_5$ are critical for correct extraction of EDM values on a lattice.
This is evident from the following example: the nucleon EDFF can be extracted from the timelike
component of the vector current, which has the following $\CP$-odd correction to the matrix
element between nucleon states polarized in the $\hat{i}$-th direction,
\begin{equation}
\label{eqn:edm_from_J4}
\langle \vec p^\prime=0 | V_4|\vec p=-\vec q\rangle_{\CPviol} \propto \frac{q_i}{m} \big[
  (1+\tau) F_3(Q^2) + \alpha_5 G_E(Q^2) \big]\,,
\end{equation}
with $\tau = Q^2 / (4 M_N^2)$ and $G_E$ the Sachs electric form factor.
For the proton with nonzero charge $Q=G_{Ep}(0)=1$, a biased value of $\alpha_5$ will lead
to an incorrect lattice EDM value.
For the neutron with $G_{En}(0)=0$, this may be less problematic.
However, the nonzero contribution of $G_{En}$ for $Q^2>0$ can make extraction of the neutron EDM
from the $Q^2\to0$ extrapolation more complicated (e.g., require more sophisticated $Q^2$ fits)
as well as affect the $Q^2$-dependence of the EDFF and the neutron's Schiff moment
$F^\prime_{3n}(0)$.

\begin{table}[htb]
\centering
\caption{
  Gauge ensembles used in this study. 
  The second column shows the action used and the reference where the ensemble was analyzed.
  \label{tab:gauge_ens}}
\begin{tabular}{ll|ccc|rr}
\hline\hline
$L_x^3\times L_t\times L_5$ & $S_F$[Ref] &
  $a\text{ [fm]}$ &   
  $m_\pi\,[\mathrm{MeV}]$ & $m_N\,[\mathrm{GeV}]$ &  
  Conf & Obsv.\\
\hline
$24^3\times64\times16$ & DWF\cite{Aoki:2010dy} &
  0.1105(6) & 340(2) & 1.178(10) & 
  1400 & 
  $\theta$-nEDM \\
$48^3\times96\times24$ & MDWF\cite{Blum:2014tka} & 
  0.1141(3) & 139.2(4) & 0.945(6) & 
  130 &  $\mcP$,$\mcC$-nEDM \\
\hline\hline
\end{tabular}
\end{table}

In this study, we use ensembles of QCD gauge configurations generated by the RBC/UKQCD
collaboration employing Iwasaki gauge action and $N_f=2+1$ dynamical chiral-symmetric
fermions with (M\"obius) domain wall action (see Tab.~\ref{tab:gauge_ens}).
One ensemble has unphysical heavy pion mass $m_\pi\approx340\text{ MeV}$ and is used to
study the $\theta_\text{QCD}$-induced nEDM.
The reason for using a heavy pion mass is that the effect of $\theta_\text{QCD}$ term is reduced
at lighter quark masses (and vanishes in the chiral limit), therefore physical light-quark
calculations would be extremely challenging. 
An estimate is provided in the next section.
The other ensemble is generated with a physical pion mass $m_\pi\approx139\text{ MeV}$ 
and is used to calculate nucleon form factors and nEDMs induced by quark-gluon 
chromo-EDMs in QCD with realistic parameters.

\section{Nucleon EDM induced by the $\theta_\text{QCD}$-term}

Studying $\theta_\text{QCD}$-induced nEDM is complicated by the statistical noise due to
the global nature of the topological charge~(\ref{eqn:Qtopo}).
Its fluctuation $(\delta Q)^2=\langle Q^2\rangle \propto V_4$ grows with the lattice
volume $V_4$ and leads to large statistical uncertainty in $\CP$-odd correlation
functions~(\ref{eqn:twopt_cpviol},\ref{eqn:threept_cpviol}).
As suggested in Refs.~\cite{Shintani:2015vsx,Liu:2017man}, contributions to $Q$
from distant sites may be neglected in computing nEDM.
However, spatial restriction of $Q$ may bias EDM results, 
for example if the ``effective'' parity mixing angle $\alpha_5$ is different in the
nucleon~(\ref{eqn:twopt_cpviol}) and the nucleon-current~(\ref{eqn:threept_cpviol}) correlation
functions, as indicated by Eq.~(\ref{eqn:edm_from_J4}).
Such difference may be produced by non-identical spatial or timelike restriction of the partial
topological charge in these $\CP$-odd Green's functions, which results in nucleon interpolating 
operators acting on vacua with different amount of $\CP$ violation.
To illustrate this point, 
consider the $\CPviol$ interaction 
that is turned on at some moment $t<0$.
The QCD vacuum takes some Euclidean time $\Delta t$ to evolve into the new $\CP$-violating state
$|\vac\rangle\to|\vac\rangle_{\CPviol}$.
Nucleon operators $\bar{N}$ acting on such transient vacuum state will have time-dependent
overlap $\langle\tilde{n}|\bar{N}|\vac(t)\rangle$ with the new nucleon-like states 
$|\tilde{N}^{(\pm)}\rangle = |N^{(\pm)}\rangle \pm i\alfive |N^{(\mp)}\rangle$
leading to ambiguity in the extracted values of the parity-mixing angle $\alpha_5$ and EDFF $F_3$.
A similar argument applies to the nucleon sinks.

\begin{figure}[htb]
\begin{minipage}{.4\textwidth}
  \centering
  \includegraphics[width=.65\textwidth]{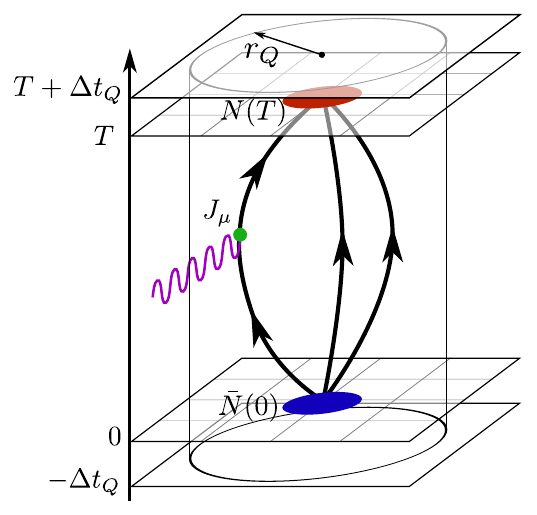}
\end{minipage}~\hspace{.05\textwidth}~
\begin{minipage}{.5\textwidth}
  \caption{
  Constrained sampling of the topological charge density~(\ref{eqn:Qtopo_cuts})
  for reducing the statistical noise in the
  $\CP$-odd three-point correlation functions~(\ref{eqn:threept_cpviol}), 
  as well as the $\CP$-odd two-point correlation functions~(\ref{eqn:twopt_cpviol}).
  \label{fig:qtopo_reduced}}
\end{minipage}
\end{figure}

To avoid this ambiguity, in our study we restrict the topological charge estimator separately in
time and space to a cylindrical volume $V_Q$ (Fig.~\ref{fig:qtopo_reduced}),
\begin{equation}
\label{eqn:Qtopo_cuts}
\tilde{Q}(\Delta t_Q,r_Q) 
  = \frac1{16\pi^2} \sum_{x\in V_Q}\Tr\big[ \hat{G}_{\mu\nu} \tilde{\hat{G}}_{\mu\nu}\big]_{x}\,,
\quad (\vec x,t)\in V_Q :\left\{\begin{array}{l} 
  |\vec x-\vec x_0|\le r_Q\,, \\
  t_0 - \Delta t_Q < t < t_0 + \tsep + \Delta t_Q\,,
\end{array}\right.
\end{equation}
where $t_0$ is the location of the nucleon source and $t_0+\tsep$ is the location of the nucleon
sink.
The $\CP$-odd correlation functions~(\ref{eqn:twopt_cpviol},\ref{eqn:threept_cpviol}) are
computed entirely inside the region~(\ref{eqn:Qtopo_cuts}) where $\CP$ violation is present
(i.e. where the reduced topological charge $\tilde{Q}$ is sampled).
The timelike cuts applied to $\tilde{Q}$ are symmetric with respect to the nucleon
sources and sinks and equal in the nucleon~(\ref{eqn:twopt_cpviol}) and
nucleon-current~(\ref{eqn:threept_cpviol}) correlation functions.
Additionally, we restrict $\tilde{Q}$ sampling in space to a 3D ball centered 
on the nucleon source, to further reduce the stochastic noise on large-volume lattices.
However, this restriction may interfere with the momentum projection 
in Eq.(\ref{eqn:threept_cpviol}) that requires summation over all $\vec y$ and $\vec z$.
\emph{We emphasize that convergence with $r_Q$ must be verified at
each momenta combination $p^\prime$ and $q$ to avoid bias}, especially in computing the
$Q^2$-dependence of the EDFFs and the Schiff moments.

\begin{figure}[htb]
\centering
\includegraphics[width=.49\textwidth]{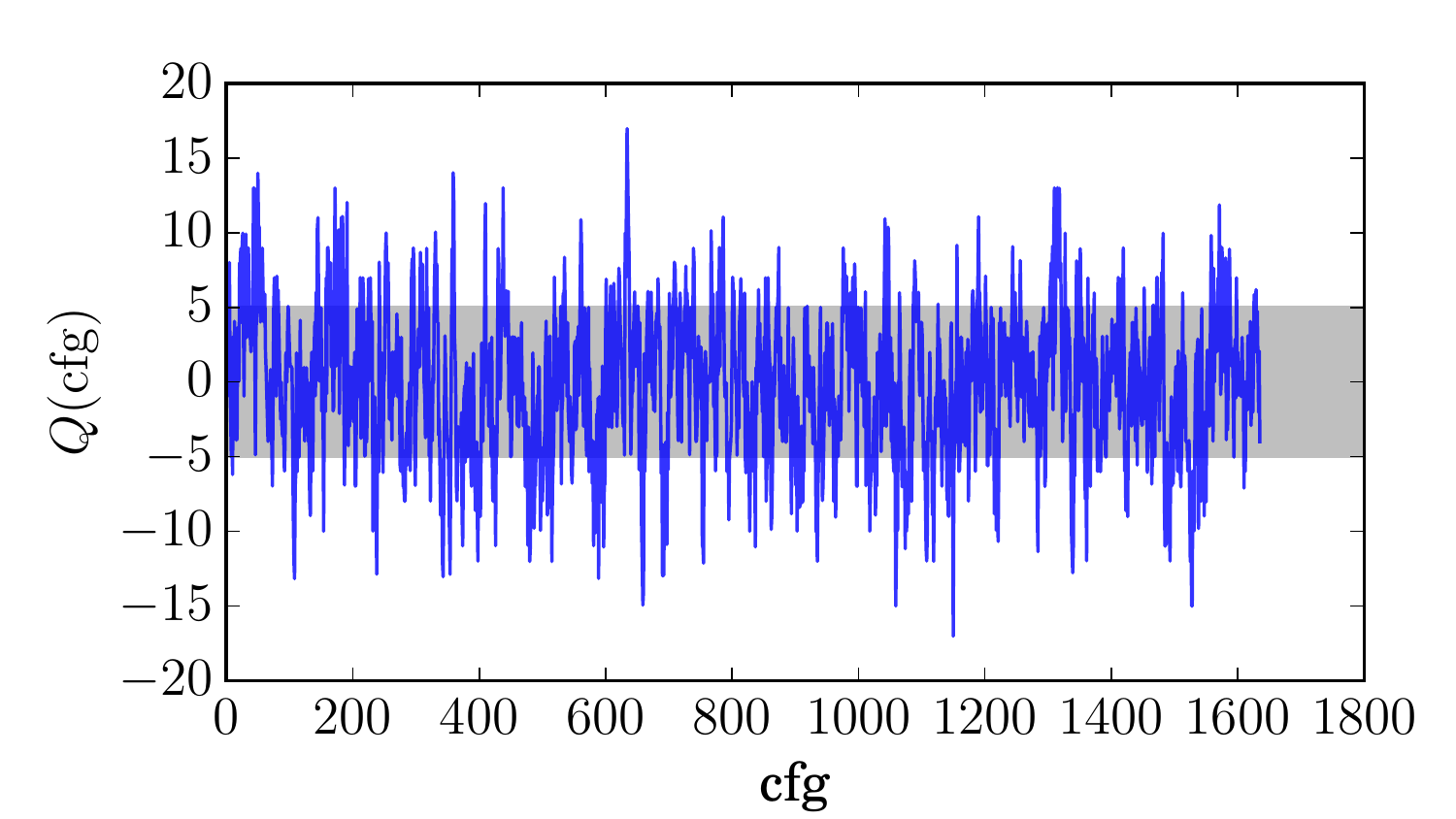}~
\hspace{.02\textwidth}~
\includegraphics[width=.49\textwidth]{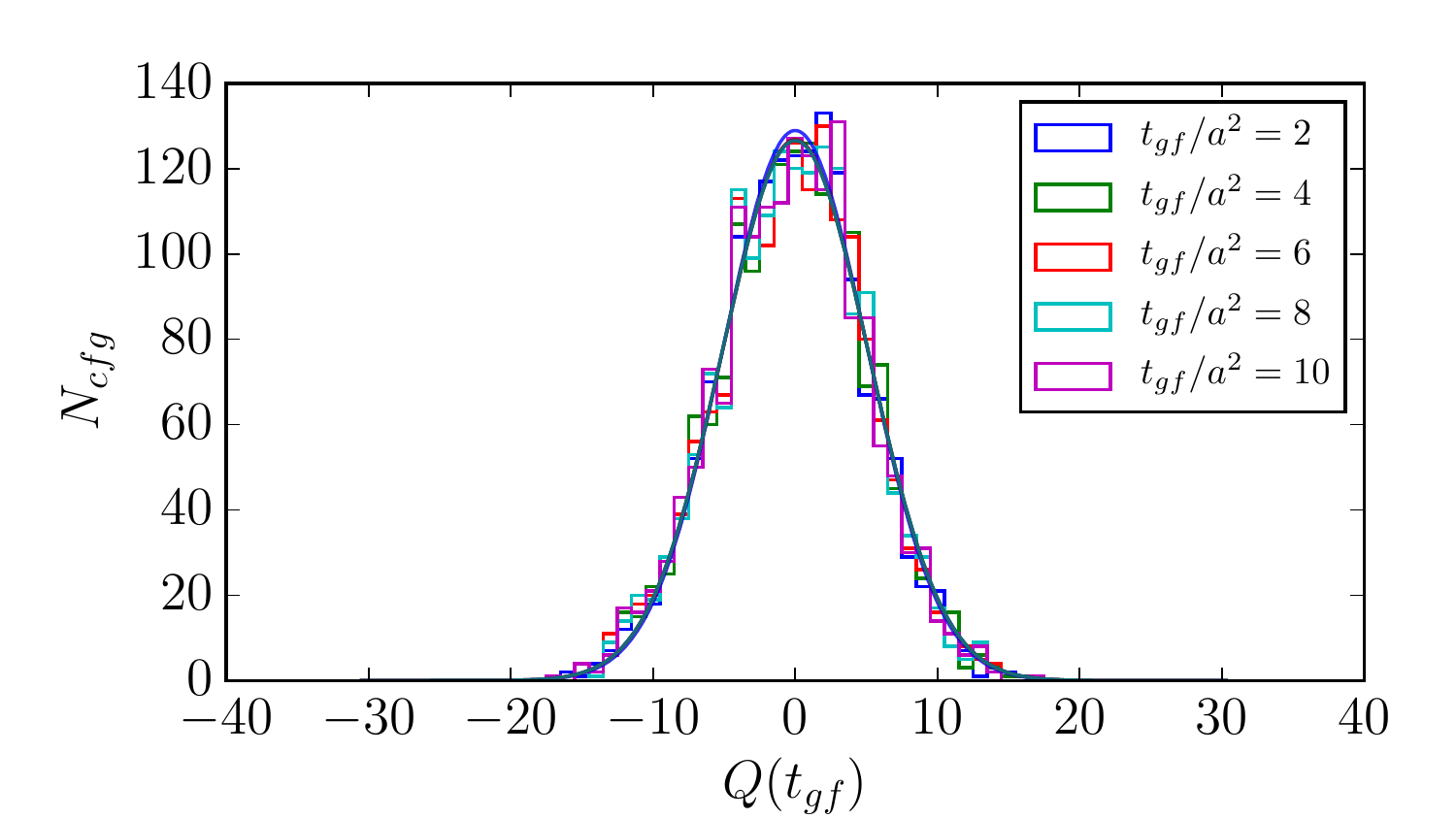}\\
\caption{
The Monte Carlo evolution history (left) and the distribution 
(right) of the global topological charge.
The topological  charge is computed with ``5-loop-improved'' field strength 
tensor~\cite{deForcrand:1997esx} using gradient-flowed gauge links (only $t_{gf}/a^2=8$ shown on the left).
Also shown on the right are the Gaussian distributions with the corresponding values of 
$\langle Q^2\rangle$ for a range of $t_{gf}/a^2$.
}
\label{fig:theta_evol_hist}
\end{figure}

We use the lattice QCD ensemble with unphysical heavy pion mass $m_\pi\approx340\text{ MeV}$
(see Tab.~\ref{tab:gauge_ens}) to enhance the nEDM value in this preliminary study, since the
$\theta_\text{QCD}$-induced EDM decreases with decreasing $m_q\propto m_\pi^2$.
We calculate 64 low-precision and 1 high-precision samples using the \emph{AMA} 
sampling method~\cite{Shintani:2014vja}.
We analyze 1,400 gauge configurations separated by 5 MD steps to obtain 89,600 samples; samples
from each 10 MD steps (2 adjacent gauge configurations) are binned together.
The topological charge density in Eq.~(\ref{eqn:Qtopo_cuts}) is calculated from 
``5-loop-improved'' field strength tensor $\hat{G}_{\mu\nu}$~\cite{deForcrand:1997esx}
computed from gradient-flowed~\cite{Luscher:2010iy,Luscher:2011bx,Luscher:2013vga} gauge
fields($\tau_{GF}=8a^2$).
The Monte Carlo evolution history of the total topological charge and its distribution are shown
in Fig.~\ref{fig:theta_evol_hist}.

\begin{figure}[htb]
\centering
\includegraphics[width=\textwidth]{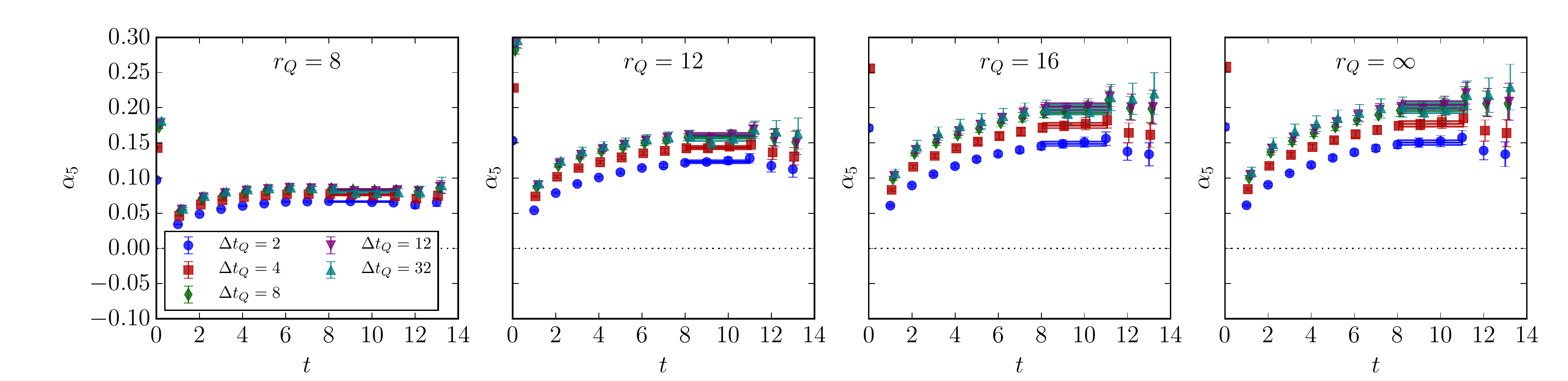}
\caption{
  The nucleon parity mixing angle~(\ref{eqn:alfive_ratio}) and its dependence on the spatial and
  temporal cuts in the reduced topological charge $\tilde{Q}(\Delta t_Q, r_Q)$.
  \label{fig:parmixing_dwf24c64}}
\end{figure}

First we study the effect of reduced topological charge sampling on the mixing angle $\alpha_5$.
The mixing angle $\alpha_5$ is estimated with the $\{t,\Delta t_Q,r_Q\}$-dependent ratio
\begin{equation}
\label{eqn:alfive_ratio}
\alfivehat^{eff}(t) 
  = -\frac{\Tr\big[ T^+\gamma_5 \, \delta_{\tilde{Q}(\Delta t_Q,r_Q)}^\CPbar C_{N\bar N} (t)\big]}
          {\Tr\big[ T^+ \, C_{N\bar N}(t)\big]}
  \stackrel{t\to\infty} = \frac{\alfive}{\theta} \,.
\end{equation}
where $T^+=\frac{1+\gamma_4}2$ is the positive-parity projector.
Results for different values of $\Delta t_Q,\,r_Q$ are shown in
Fig.~\ref{fig:parmixing_dwf24c64}.
We generally observe convergence to the results obtained with the full topological charge 
Q~(\ref{eqn:Qtopo}) for $\Delta t_Q\gtrsim 8a$.
However, for the spatial cut $r_Q$ there is no convergence up to $r_Q\approx12a$, which is
$\approx52\%$ of the spatial volume.
We conclude that the lattice volume $V_3=(24a)^3\approx(2.7\text{ fm})^3$ 
is insufficient to benefit from the spatial cut $r_Q$, and should be explored with larger
spatial volumes.

\begin{figure}[htb]
\centering
\includegraphics[width=\textwidth]{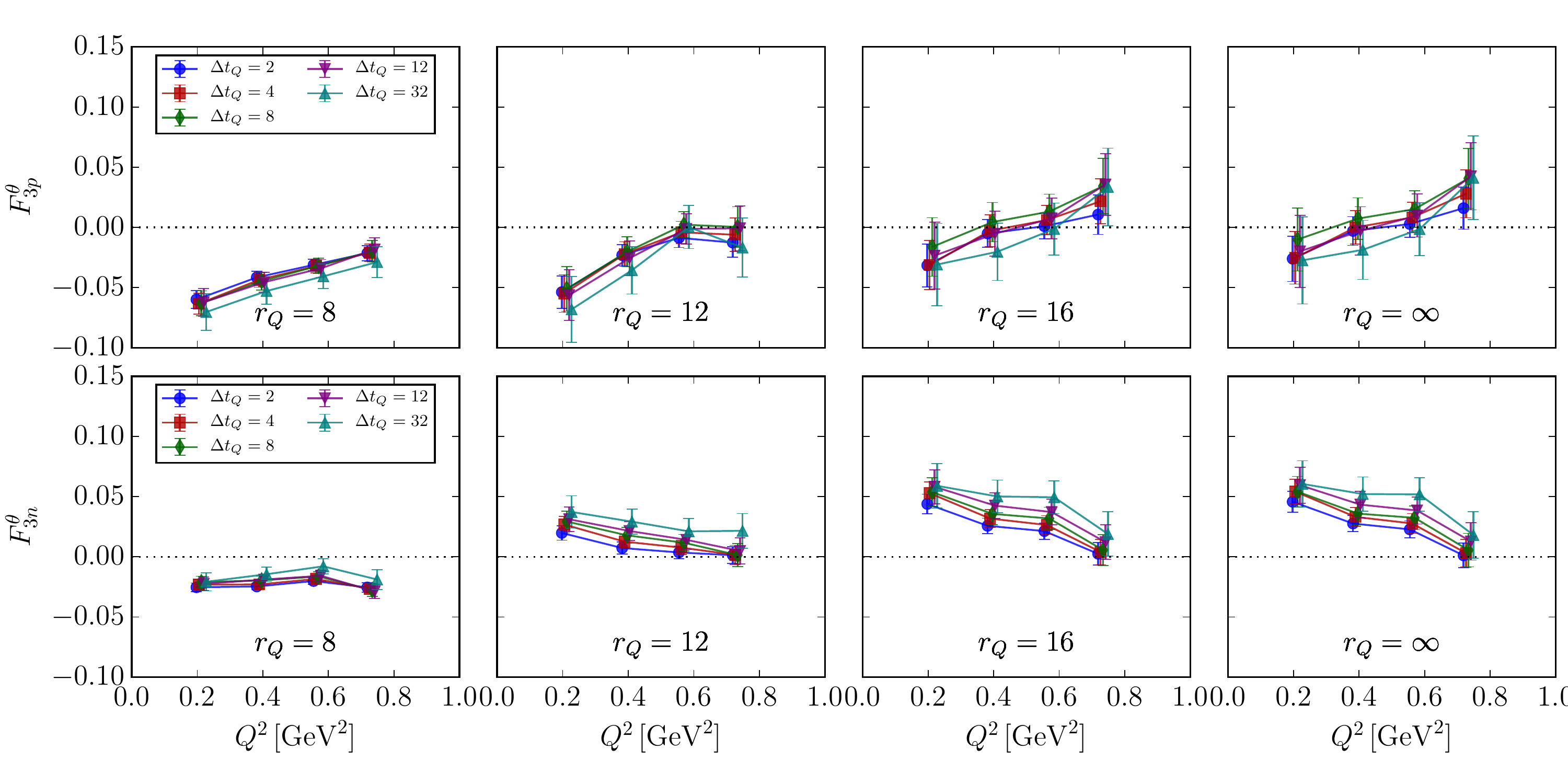}
\caption{
  Proton and neutron electric dipole form factors induced by the $\theta_\text{QCD}$-term 
  from lattice calculations with $m_\pi\approx340\text{ MeV}$ (only quark-connected contractions)
  and their dependence on the spatial and temporal cuts in the reduced topological charge 
  $(\Delta t_Q,r_Q)$.
  \label{fig:edff_dwf24c64_mpi330}}
\end{figure}

The neutron and proton electric dipole form factors $\hat{F}^\theta_{3n,p}=F^\theta_{3n,p}/\theta$ 
computed for a range of $\Delta t_Q,r_Q$ values are shown in Fig.\ref{fig:edff_dwf24c64_mpi330}. 
We compute only connected diagrams in this study.
The values for $\hat{F}^\theta_3$ are obtained using Eq.~(\ref{eqn:edm_from_J4}) with one value of
source-sink separation $\tsep=8a$.
Similarly to $\alfive$, we observe convergence for $\Delta t_Q\gtrsim8a$ but lack of convergence
for $r_Q\lesssim12a$.
Most importantly, we observe statistically significant value for the neutron $F_3$ even with the
full value of the topological charge $Q$, which has no bias from reduced sampling
$Q\to\tilde{Q}(\Delta t_Q, r_Q)$.
We can make \emph{a very preliminary ``ballpark'' estimate} for the value of 
$\hat{F}^\theta_{3n}(0)\approx0.05$ at this heavy pion mass, \emph{which should be taken with a
$100\%$ uncertainty} since it does not take into account excited state effects or extrapolation
$Q^2\to0$.
This value should only be used to check consistency with phenomenology and earlier lattice
QCD calculations.
For example, the \emph{corrected} value from calculations with Wilson
fermions~\cite{Guo:2015tla} constrains $|\hat F^\theta_3(0)|\lesssim0.06$ at a close value of
the pion mass $m_\pi\approx360\text{ MeV}$.
Leading-order extrapolation~\cite{Crewther:1979pi,Hockings:2005cn} 
$\hat d^\theta_n\propto m_{u,d}\propto m_\pi^2$ to the physical point yields values
\begin{equation}
\label{eqn:F3n_phys_estimate}
|\hat F_{3n}^{\theta,\text{phys}}|\approx0.01\,,
\quad\text{ or }
|\hat d^{\theta,\text{phys}}_{n}|
  = \frac{e}{2m_N} |\hat F_{3n}^{\theta,\text{phys}}|
  \approx0.001\,e\cdot\mathrm{fm}
\,,
\end{equation}
which is consistent with estimates from ChPT
and the QCD sum rules~\cite{Engel:2013lsa}.

\begin{figure}[htb]
\centering
\includegraphics[width=\textwidth]{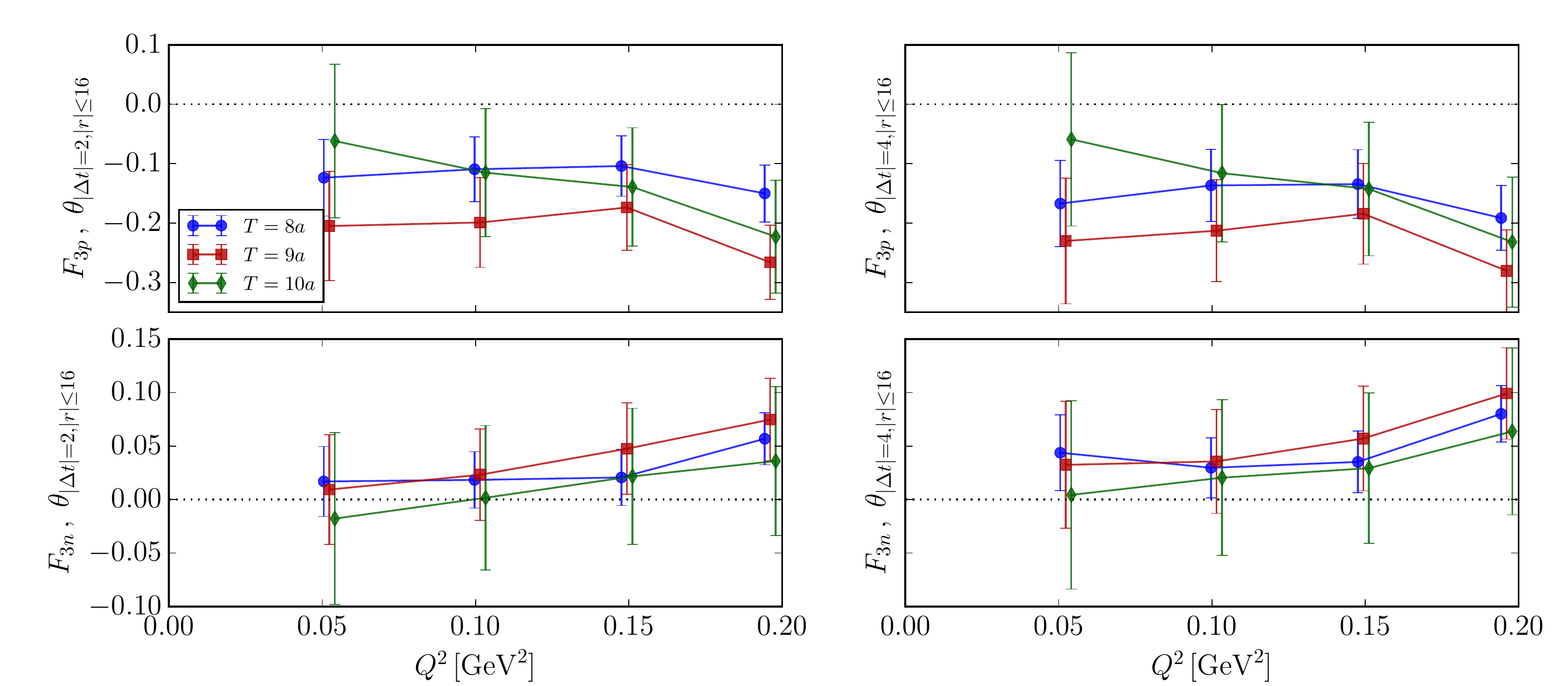}
\caption{
Proton and neutron electric dipole form factors induced by $\theta_\text{QCD}$-term 
from lattice calculations with physical quark masses.
}
\label{fig:edff_dwf48c96_mpiphys}
\end{figure}

Using our rough estimate for $\hat F^\theta_{3n}$, we can project the effort required for
computing nEDM at the physical point, which is required to avoid model dependence due to
pion mass extrapolations $m_\pi \to m_\pi^\text{phys}$.
We have performed initial calculations using physical-quark ensembles with
$m_\pi\approx139\text{ MeV}$ (see Tab.~\ref{tab:gauge_ens}) with $\approx$ 33,000 statistical
samples and very aggressive time and space cuts in the topological charge estimator
$\tilde{Q}(\Delta t=2a,r_Q=16a)$.
We observe no signal for the neutron EDFFs (see Fig.~\ref{fig:edff_dwf48c96_mpiphys}), 
and the results are consistent with zero with the statistical uncertainty 
$\delta F_{3n}\approx0.05\ldots0.10$.
In comparison to the estimate~(\ref{eqn:F3n_phys_estimate}) above, we expect that the current
signal-to-noise ratio (SNR) $\approx0.01/0.05=0.2$ has to be improved at least by a factor of
5-10, which requires $\times(25\ldots100)$ more statistics.
Alternative computing methods may have to be employed such as dynamical (imaginary)
$\theta^I$-term first explored in Ref.~\cite{Aoki:2008gv}.
Because nEDM calculations depend on contributions from non-trivial topological sectors,
dynamical $\theta^I$-term improves importance sampling for the EDM signal by inducing
nonzero average topological charge $\langle Q\rangle\ne0$.
The dynamical $\theta^I$-term becomes more important at lighter pion masses, where 
light quarks suppress the fluctuation of the topological charge.

\section{Nucleon EDM induced by quark chromo-EDM}

In this section, we report results from the ongoing calculations of nucleon EDM induced by the
dimension-5(6)\footnote{
  Quark-gluon chromo-EDM operator has dimension 6 above the electroweak scale 
  due to the Higgs field factor required by the electroweak symmetry.
}
chromo-electric quark-gluon interaction~(\ref{eqn:cedm_def}). 
We use the physical point ensemble (see Tab.~\ref{tab:gauge_ens}) and evaluate 256 low-precision
and 4 high-precision samples on each of 130 statistically-independent gauge configurations
separated by 40 MD steps, for the total of 33,280 statistical samples. 
All samples from the same gauge configuration are binned together.

\begin{figure}[htb]
\centering
\includegraphics[width=.49\textwidth]{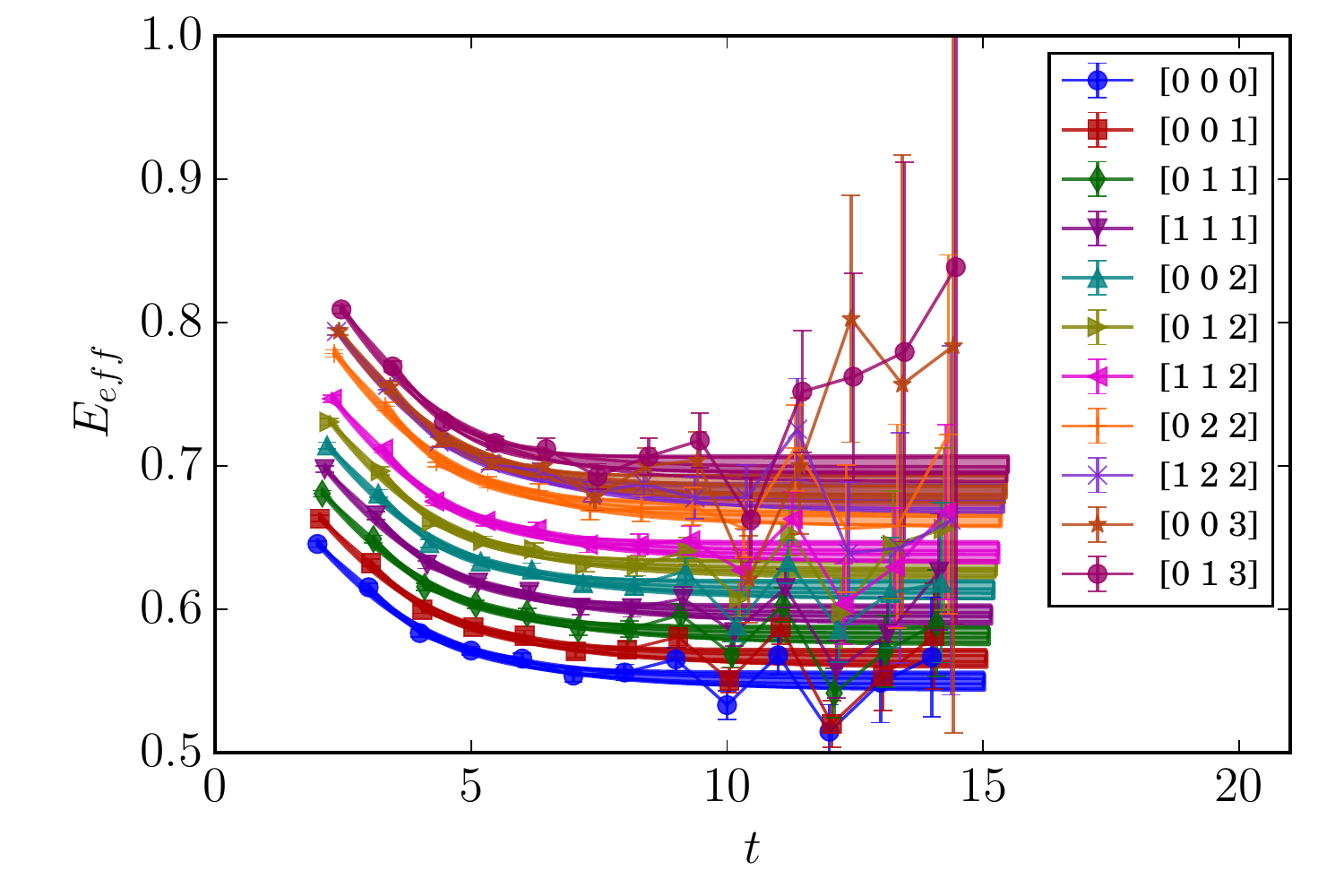}~
\hspace{.02\textwidth}~
\includegraphics[width=.49\textwidth]{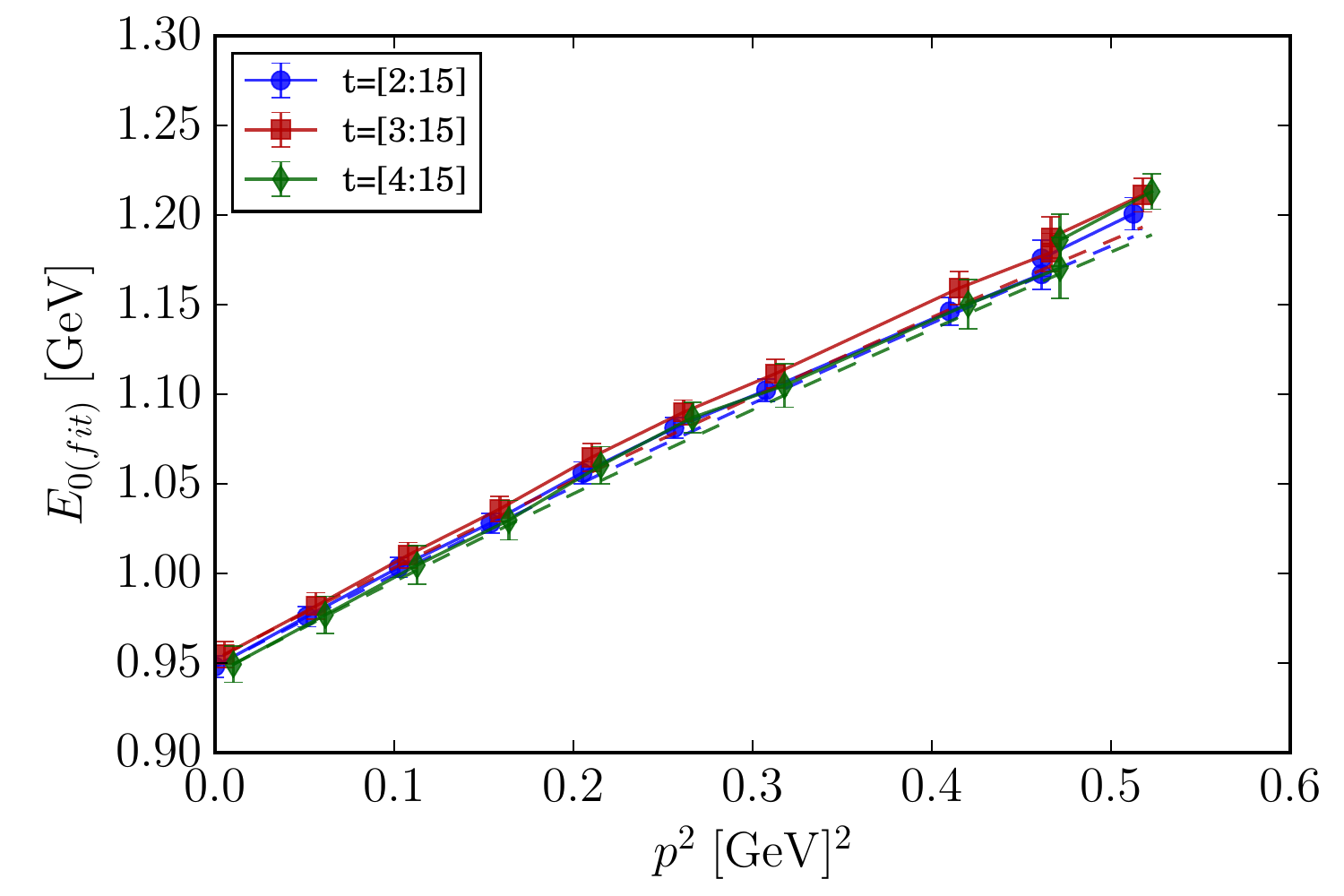}
\caption{
(Left) Nucleon masses and energies computed with 2-state fits.
Results from fits in three ranges with $t_\text{min}/a=2,3,4$ are shown with shaded bands and are 
consistent for each lattice momentum.
(Right) Nucleon energy vs. momentum in physical units. 
The dashed lines show the continuum dispersion relation that uses the lattice mass value 
$m_N=E_N(\vec p=0)$.
\label{fig:ff_mN_fit_disp}
}
\end{figure}

The results for the nucleon mass and energies with momenta up to $|\vec p|\approx0.72\,\text{GeV}$
are shown in Fig.~\ref{fig:ff_mN_fit_disp}.
These energies are extracted with unconstrained 2-state fits for a number of fit ranges starting 
at $t_\text{min}/a=2,3,4$, and are all consistent with each other.
The momentum dependence of the nucleon energy (Fig.~\ref{fig:ff_mN_fit_disp}, right) is
compared to the continuum dispersion relation, and its close agreement indicates that the
discretization errors are small.

\begin{figure}[htb]
\centering
\includegraphics[width=\textwidth]{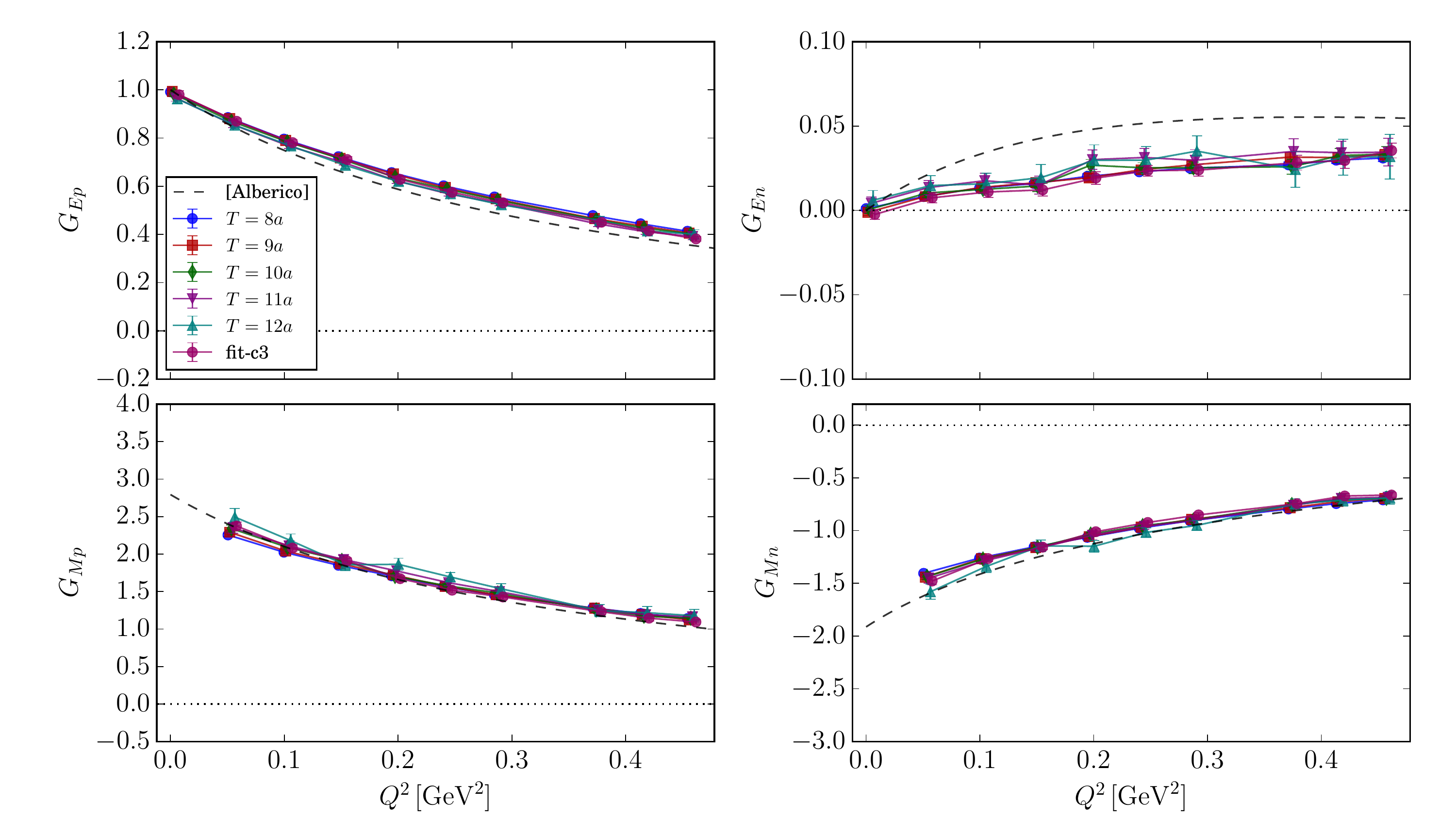}
\caption{
Nucleon electromagnetic form factors (connected contributions)
from lattice calculations with physical quark masses. 
\label{fig:ff_mpiphys}
}
\end{figure}

In Figure~\ref{fig:ff_mpiphys}, we show preliminary results for the proton and neutron
electromagnetic Sachs form factors $G_{E,M\,p,n}(Q^2)$ and their comparison with the
phenomenological fits to experimental data~\cite{Alberico:2008sz}.
These form factors are extracted using the standard ``ratio'' method (see, e.g.,
Ref.~\cite{Hagler:2007xi}) for five fixed source-sink separations $\tsep=(8\ldots12)a$ as well as
2-state fits using the state energies obtained from the nucleon two-point correlation functions.
Magnetic form factors $G_{Mp,n}$ show reasonable agreement with phenomenology. 
However, the electric form factors $G_{Ep,n}$ disagree for both the proton and the neutron,
which may be attributed to the missing contribution from the disconnected contractions, which
are not currently evaluated.

\begin{figure}[htb]
\centering
\includegraphics[width=.8\textwidth]{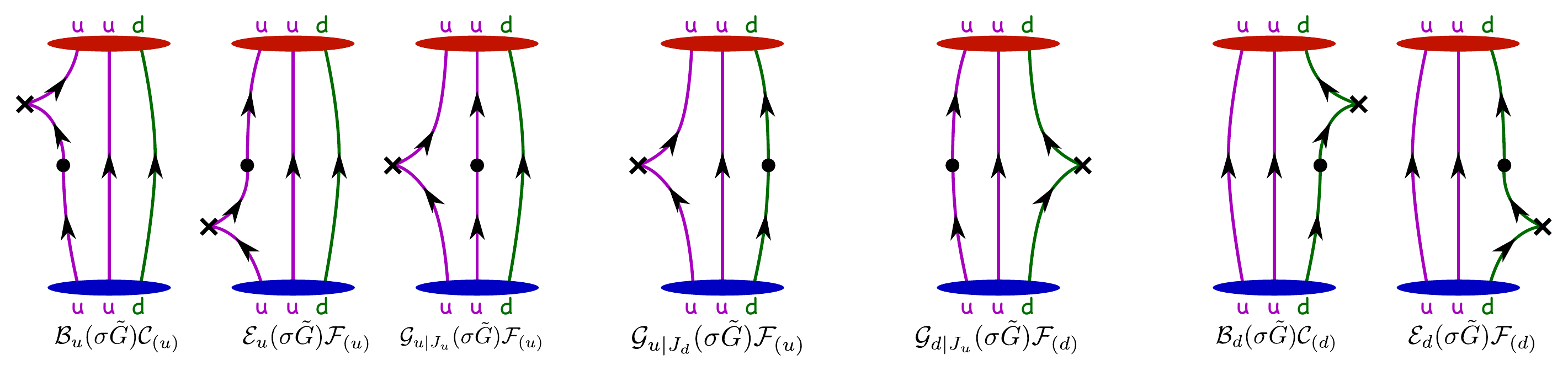}\\
\includegraphics[width=.57\textwidth]{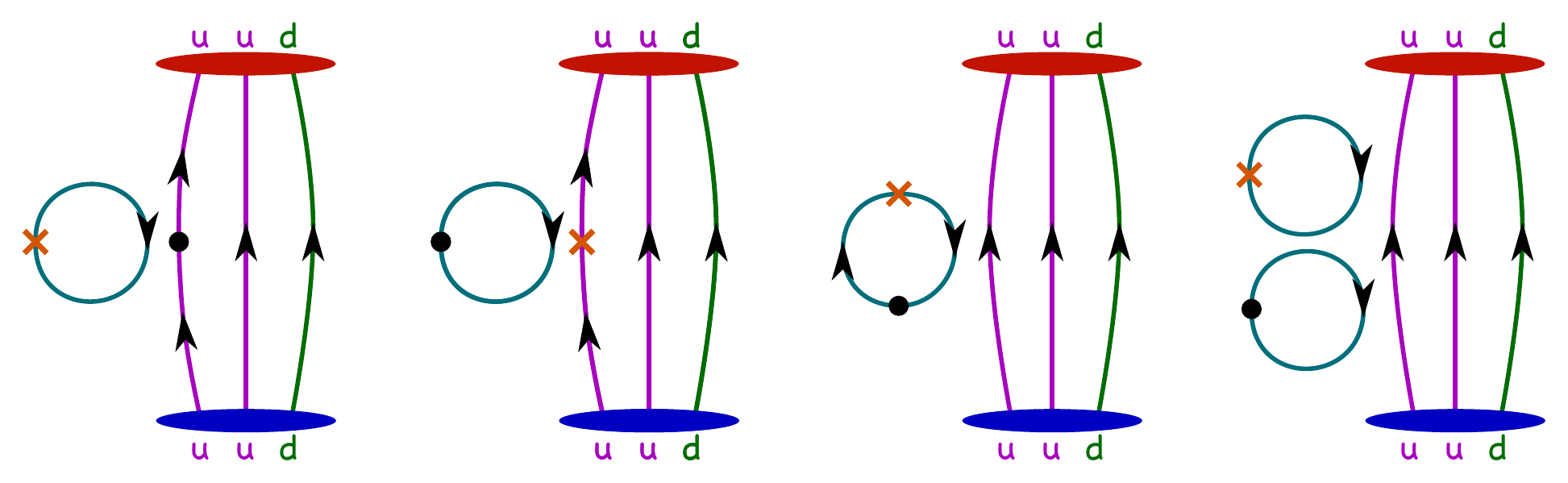}
\caption{
Connected (top) and disconnected (bottom) lattice contractions for computing nucleon electric
dipole form factors induced by chromo-EDM.
}
\label{fig:edff_diags}
\end{figure}

Calculation of cEDM-induced nucleon EDM require insertions of quark bilinear
operators~(\ref{eqn:cedm_def}), separately for each flavor.
The lattice gluon field strength $\hat{G}_{\mu\nu}$ in the chromo-EDM
density~(\ref{eqn:cedm_def}) on a is computed with the symmetric (``clover'') operator using
only square $1\times 1$ plaquettes,
\begin{equation}
\label{eqn:Gmunu_clover}
\begin{aligned}
\big[\hat{G}_{\mu\nu}\big]_x^\text{clov} &= \frac1{8i} \big[
  ( U^P_{x,+\hat\mu,+\hat\nu} + U^P_{x,+\hat\nu,-\hat\mu} 
  + U^P_{x,-\hat\mu,-\hat\nu} + U^P_{x,-\hat\nu,+\hat\mu}) 
  - \mathrm{h.c.}\big]\,.
\end{aligned}
\end{equation}
We evaluate only fully connected diagrams for both the $\CP$-even and $\CP$-odd correlation
functions~(\ref{eqn:twopt_cpviol},\ref{eqn:threept_cpviol}), which are shown in
Fig.\ref{fig:edff_diags}(top) for the latter.
Disconnected diagrams (see Fig.\ref{fig:edff_diags}, bottom), which are required for a complete
unbiased calculation of isoscalar EDMs and effects of isoscalar quark chromo-EDMs, are more
computationally demanding and will be evaluated in the future.

\begin{figure}[htb]
\centering
\includegraphics[width=\textwidth]{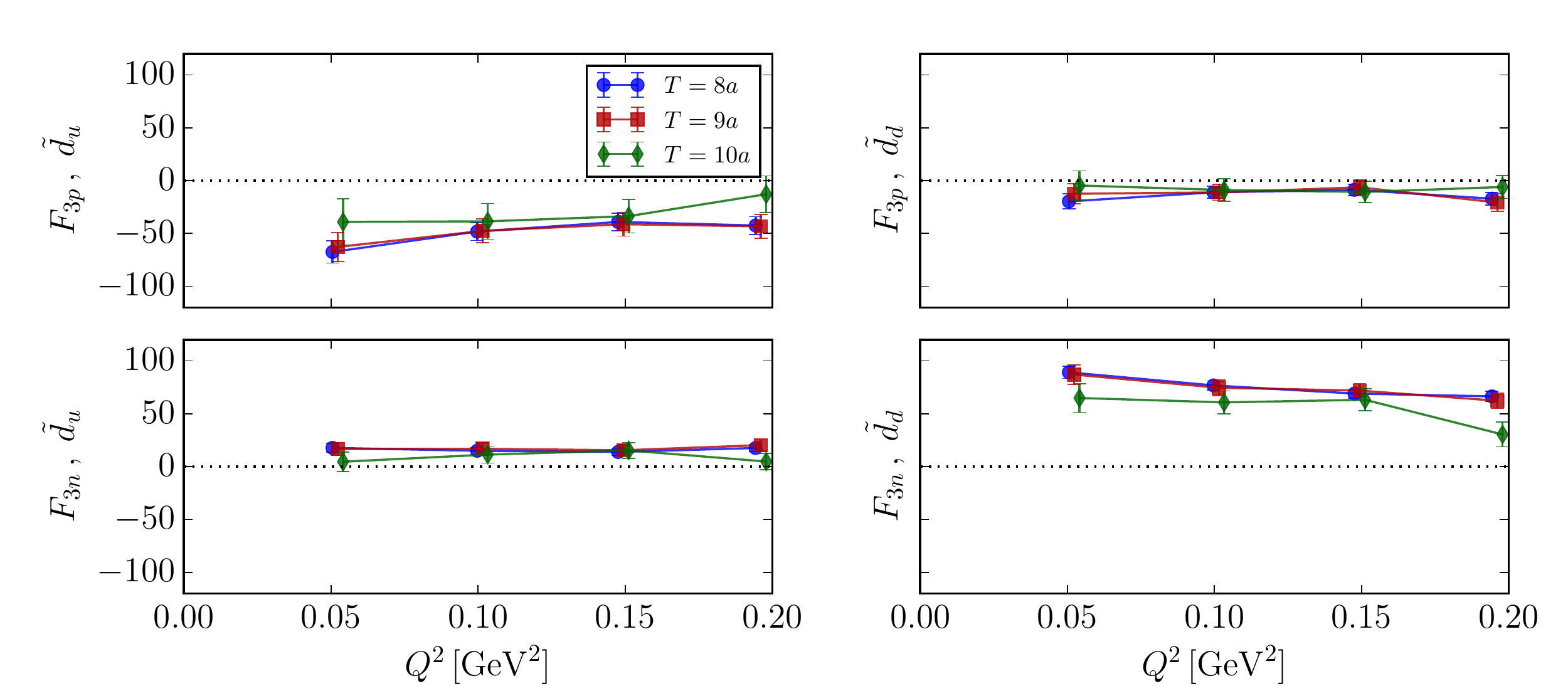}
\caption{
Proton and neutron electric dipole form factors induced by (lattice bare) chromo-EDM
from lattice calculations with physical quark masses. 
Only connected contributions.
}
\label{fig:edff_cedm_mpiphys}
\end{figure}

\begin{figure}[htb]
\centering
\includegraphics[width=\textwidth]{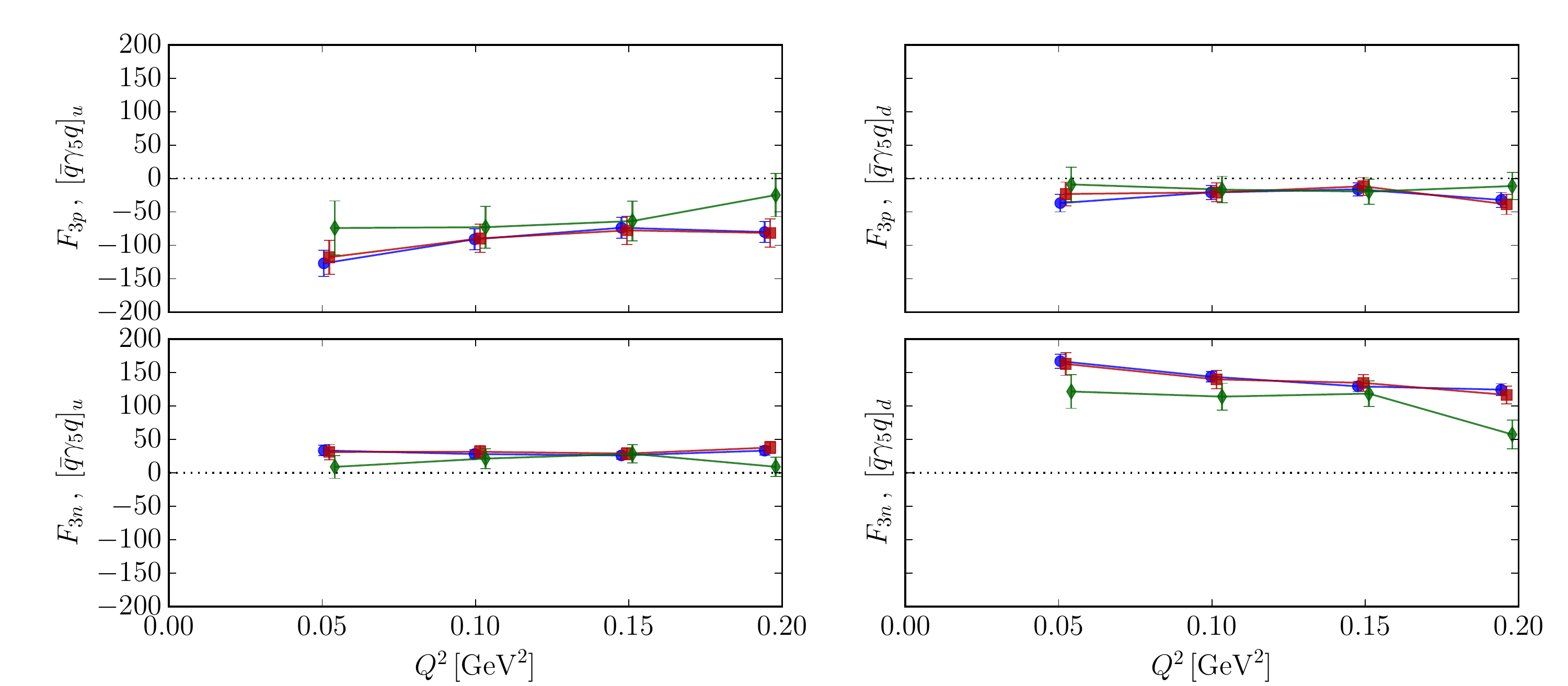}
\caption{
Proton and neutron electric dipole form factors induced by (lattice bare) quark pseudoscalar
density from lattice calculations with physical quark masses.
Only connected contributions.
}
\label{fig:edff_psc_mpiphys}
\end{figure}

In Figure~\ref{fig:edff_cedm_mpiphys} we show proton and neutron
EDFF induced by the \emph{unrenormalized (bare lattice)} quark chromo-EDM $\mcC$.
These form factors are extracted using the ``ratio'' method with fixed source-sink separations
$\tsep=(8\ldots10)a$.
Data shows signal for both $\mcC$ and $\mcP$.
There is a peculiar dependence of nEDM on the flavor structure of $\CP$ violation:
the proton and the neutron EDMs are induced by the $\CP$ violation in the ``unpaired'' flavors,
i.e. in $u$- and $d$-quarks, respectively.
Finally, in Figure~\ref{fig:edff_psc_mpiphys} we show the nucleon EDM induced by the quark
pseudoscalar density that is required for renormalizing and mixing subtraction for results shown
in Fig.~\ref{fig:edff_cedm_mpiphys}.

The final results for the chromo-EDM-induced nucleon EDMs require renormalization, that has to
be computed nonperturbatively on a lattice.
One proposed scheme is RI-SMOM, and perturbative matching to the $\MSbar$ scheme has been
calculated~\cite{Bhattacharya:2015rsa}.
Another approach is the position-space scheme~\cite{Gimenez:2004me,Chetyrkin:2010dx},
calculations of perturbative matching for which are underway.

\section{Summary and Outlook}
Calculations of nEDM on a lattice are important for interpreting constraints or results from 
nucleon and nuclei EDM measurements.
Ongoing calculations of nEDM induced by dim-5(6) quark-gluon $\CP$ violation show promising
results at the physical point.
However, their final precision will depend on renormalization that has not been computed yet,
and renormalized results may require substantially more statistics.
In contrast, calculations of $\theta_\text{QCD}$-induced nEDM at the physical point 
will be challenging and will require special techniques to tame the statistical noise caused by 
fluctuations of the global topological charge.
Direct calculations at the physical point may be at the limit of the current computing
capabilities, and one may have to use ChPT extrapolations of unphysical heavy-pion results.
Another approach is to simulate QCD with dynamical $\theta^I_\text{QCD}$ term to enhance
importance sampling for the $\CPviol$ observables.

\section*{ACKNOWLEDGEMENTS}
We are grateful for the gauge configurations provided by the RBC/UKQCD collaboration.
This research used resources of the Argonne Leadership Computing Facility, which is a DOE Office
of Science User Facility supported under Contract DE-AC02-06CH11357,
and Hokusai supercomputer of the RIKEN ACCC facility.
SS is supported by the RHIC Physics Fellow Program of the RIKEN BNL Research
Center.
TI is supported in part by US DOE Contract DESC0012704(BNL), and JSPS KAKENHI grant numbers
JP26400261, JP17H02906.
HO is supported in part by JSPS KAKENHI Grant Numbers 17K14309.

\bibliographystyle{aip}
\bibliography{bib}
 
\end{document}